\documentclass[acmsmall]{acmart}
\usepackage{algorithm}
\usepackage[noend]{algpseudocode}
\usepackage{enumitem}
\usepackage{xcolor}

\newif\ifshowrevisions
\ifshowrevisions
  \newcommand{\rev}[1]{{\color{blue}#1}}
\else
  \newcommand{\rev}[1]{#1}
\fi
\AtBeginDocument{%
  }

\begin{document}

\newcommand{\sysname}{\textsc{Rain}}
\title{\sysname{}: RDMA-assisted In-Network Scheduling for Microsecond-scale Workloads}

\setcopyright{cc}
\setcctype{by}
\acmJournal{PACMNET}
\acmYear{2026} \acmVolume{4} \acmNumber{CoNEXT2} \acmArticle{22}
\acmMonth{6} \acmDOI{10.1145/3808670}

\author{Zhihuang Ma}
\affiliation{%
  \institution{University of Science and Technology of China}
  \city{Hefei}
  \country{China}}
\email{zhma@mail.ustc.edu.cn}

\author{Xingming Cui}
\affiliation{%
  \institution{University of Science and Technology of China}
  \city{Hefei}
  \country{China}}
\email{cuixingming@mail.ustc.edu.cn}

\author{Xiaoliang Chen}
\affiliation{%
  \institution{University of Science and Technology of China}
  \city{Hefei}
  \country{China}}
\email{xlichen@ieee.org}

\author{Zuqing Zhu}
\affiliation{%
  \institution{University of Science and Technology of China}
  \city{Hefei}
  \country{China}}
\email{zqzhu@ieee.org}

\renewcommand{\shortauthors}{Zhihuang Ma, Xingming Cui, Xiaoliang Chen, and Zuqing Zhu}

\begin{abstract}

Modern data center applications increasingly require microsecond-scale service time with strict tail latency requirements, which can hardly be realized with existing in-network task schedulers due to their inherent limitations. Specifically, software-based schedulers struggle to balance throughput and latency, while switch-based designs either lack global coordination, rely on packet recirculation heavily, or only offer limited support for large tasks. In light of these restrictions of the state-of-the-arts (SOTAs), we, in this work, propose \sysname{}, an RDMA-assisted in-network scheduler built atop programmable switches that maintains centralized queues while bounding worker-local queues. \sysname{} introduces a bidirectional on-switch queuing mechanism to buffer and match tasks and worker-issued tokens directly in the switch, avoiding worker-side polling and approximating the optimal behavior of join-bounded-shortest-queue without global aggregation. A switch-driven RDMA engine pre-writes arbitrarily large tasks via one-sided \texttt{WRITE} multicasts, keeping only compact metadata on the switch. Slice-aware scheduling further localizes decisions to more homogeneous queues, reducing dispersion-induced head-of-line blocking. Moreover, our study reveals that real-world systems can diverge from theoretical predictions: shallower worker queues do not always improve tail latency. Leveraging this insight, \sysname{} incorporates an adaptive scheduling strategy to optimize worker queue depths and worker-to-slice mappings at runtime. Evaluations with the real-world application RocksDB show that \sysname{} achieves 1.75$\times$ higher throughput than the best-performing SOTA while satisfying the same tail latency requirement.

\end{abstract}



\received{December 2025}
\received[accepted]{April 2026}

\maketitle

\section{Introduction}

Modern data center (DC) applications, such as latency-critical key-value stores~\cite{redis,rocksdb} and transactional databases~\cite{wei2015fast,kalia2016fasst}, increasingly operate at microsecond timescales and require strict tail-latency service-level objectives (SLOs)~\cite{dean2013tail,barroso2017attack, Lu2015_Network, Gong2014_JLT, Liu2017_TNSM, Li2022_JLT}. However, meeting these requirements stresses existing system designs, because systems that work well for millisecond-scale tasks~\cite{verma2015large,zaharia2010spark,ousterhout2013sparrow} often fail to provide the newly-required responsiveness once being pushed to the high-throughput regime. To cope with the demands of microsecond-scale workloads, system architects have explored both scale-up and scale-out approaches. Scale-up approaches originally attempted to sustain performance when facing the slowing of Moore's Law and Dennard scaling, by adding more cores, memory, and accelerators to a single machine~\cite{demoulin2021idling,didona2019size,iyer2023achieving,ibanez2021nanopu,kaffes2019shinjuku,humphries2019mind}. Nevertheless, this will increase hardware complexity, cost, and power density. Consequently, DCs now adopt scale-out architectures that distribute workloads across a cluster of machines~\cite{zhu2020racksched,liao2025towards,udayashankar2024draconis,kogias2019r2p2}, or even across multiple clusters~\cite{yassini2024horus}, to improve scalability, elasticity, and fault tolerance \cite{Yang2024_TON, Lv2025_JSAC, Xie2025_TON, Dong2025_TNSM}.

Such a scale-out scenario places a premium on effective workload scheduling, since without careful coordination, systems are prone to head-of-line (HoL) blocking, resource contention, and elevated tail latency. Meanwhile, although scheduling strategies for scale-up systems have been extensively explored~\cite{demoulin2021idling,didona2019size,iyer2023achieving,ibanez2021nanopu,kaffes2019shinjuku,humphries2019mind}, scale-out environments have much more stringent quality-of-service (QoS) requirements on latency and scalability and thus pose a few new challenges. Specifically, cluster-level throughput is normally much higher, maintaining global and timely visibility over a cluster or multiple clusters is more difficult, and communication latency becomes a non-negligible factor. Therefore, centralized software schedulers~\cite{gog2016firmament,zaharia2010spark,cidon2017appswitch} had to struggle to achieve both high throughput and low tail latency~\cite{udayashankar2024draconis,ousterhout2013sparrow}, while distributed software schedulers~\cite{ren2015hopper,ousterhout2013sparrow,boutin2014apollo} often made suboptimal decisions due to delayed coordination and incomplete visibility on loads~\cite{kogias2019r2p2,udayashankar2024draconis}. Moreover, adding a software scheduler will typically make the forwarding paths of requests one hop longer, which can further exacerbate the aforementioned issues.

To overcome these limitations, recent efforts have begun to leverage the line-rate packet processing capability of programmable switches~\cite{intel2025tofino} to realize high-performance in-network scheduling for scale-out environments~\cite{zhu2020racksched,liao2025towards,udayashankar2024draconis,kogias2019r2p2,yassini2024horus}. Although promising, they still bear several unsolved restrictions. RackSched~\cite{zhu2020racksched} and Pallas~\cite{liao2025towards} adopted a push-based scheduling model that directly pushes tasks to workers without queuing any in the switch, but this scheme would be unfavorable to tail latency under high load dispersion~\cite{kogias2019r2p2}. Even though Pallas optimistically classified tasks to reduce dispersion, its classification would become ineffective when tasks' service time could hardly be distinguished in advance~\cite{didona2019size}. R2P2~\cite{kogias2019r2p2} introduced in-network task caching but relied heavily on packet recirculation, which \rev{reduces the switch's available forwarding bandwidth; it is also} unsuitable for the workloads whose tasks are stateful and thus require strict ordering. In our experiments, we further observed that relying on recirculation might make scheduling performance diverge noticeably from its theoretical expectations. Although Draconis~\cite{udayashankar2024draconis} supports in-network caching, its design is primarily tuned for small-size tasks, while large tasks still require an additional round-trip time (RTT) to fetch data from clients or from an external in-memory store in the cluster.

The restrictions of the state-of-the-arts (SOTAs) above motivated us to propose \sysname{}, an in-network scheduler built on programmable switches for microsecond-scale workloads, in this work. \sysname{} is designed to maintain tail latency according to SLO while remaining as work-conserving as possible to maximize throughput, by leveraging two key features: 1) efficient in-network dispatching through on-switch scheduling, and 2) continuous system tuning through adaptive scheduling.

\textbf{On-switch scheduling}~(\S\ref{sec:on-switch_scheduling}).
\sysname{} performs fine-grained scheduling directly in the data plane, by maintaining centralized queues there for both tasks and admission tokens. Specifically, client-submitted tasks and worker-issued tokens are temporarily buffered in a programmable switch on the data path, and a task is dispatched only when it is paired with a token. This design bounds each worker's local queue to a configured depth. For the tasks that are too large to be buffered in the switch, \sysname{} utilizes the remote direct memory access (RDMA) based multicast to pre-write their payloads to target servers while only storing fixed-size metadata in the switch. To mitigate workload dispersion, \sysname{} partitions both in-network queues and workers into slices so that each scheduling queue handles tasks with similar characteristics~\cite{demoulin2021idling,liao2025towards,didona2019size}. In contrast to Pallas~\cite{liao2025towards}, which relies heavily on task grouping, \sysname{} adopts slicing as a conservative auxiliary optimization while retaining centralized scheduling to enforce shallow per-worker queues.

\textbf{Adaptive scheduling}~(\S\ref{sec:adaptive_scheduling}).
In addition to static in-network scheduling, \sysname{} incorporates a runtime scheduling agent that continuously adjusts the depths of workers' local queues and worker-to-slice mappings. This adaptive mechanism enables \sysname{} to react timely to workload shifts and the worker performance changes due to co-located interference~\cite{fried2020caladan}. Our study further uncovers an often-overlooked yet important phenomenon: in real-world systems, tail-latency behavior might deviate from theoretical predictions due to practical issues, and thus simply maintaining shallower worker queues does not necessarily lead to better tail performance.

We implement \sysname{} on a Tofino switch~\cite{intel2025tofino} and evaluate it using both synthetic workloads and real-world applications. The results show that \sysname{} consistently achieves the lowest $99\%$ tail latency while fully utilizing the system's processing capacity. When being evaluated with real-world application RocksDB, \sysname{} delivers 1.75$\times$ higher throughput than a SOTA in-network scheduler~\cite{liao2025towards} under the same SLO. Moreover, our adaptive scheduling enables \sysname{} to maintain low tail latency even under dynamic workloads. In all, our major contributions are as follows:
\begin{itemize}
    \item We design a bidirectional queuing mechanism to buffer and match tasks and admission tokens in a programmable switch, removing the need for worker-side task polling as well as enabling centralized control with bounded worker queues at low overhead.
    \item We develop an RDMA-assisted technique that allows programmable switches to maintain centralized queues for tasks of arbitrary size by buffering only compact metadata.
    \item We identify and characterize a phenomenon, namely, \emph{tail latency collapse}, where reducing worker-local queue depth---a practice that is traditionally believed to be beneficial for tail latency---can instead increase tail latency under nearly-overloaded conditions.
    \item We implement a full prototype of \sysname{} and perform experimental comparisons with the SOTAs, demonstrating the advantages and effectiveness of \sysname{}.
\end{itemize}

This work does not raise any ethical issues.

\section{Background and Motivation}

\subsection{In-Network Scheduling}

Existing in-network scheduling approaches can be broadly classified into two categories based on their granularities: flow scheduling and task scheduling.

Flow-level techniques, such as ECMP and other layer-4 load-balancer~\cite{zeng2022tiara}, operate on flows and are effective at balancing the volumes of aggregated traffic under high flow-entropy conditions~\cite{gangidi2024rdma}. However, their coarse granularity makes it infeasible to distinguish individual tasks within a flow, leading to HoL blocking and persistent imbalance when service demands vary across flows.

Task scheduling encompasses both millisecond-scale and microsecond-scale regimes. Systems such as Google Borg~\cite{verma2015large}, Apache Spark~\cite{zaharia2010spark}, and Sparrow~\cite{ousterhout2013sparrow} target at the millisecond-scale tasks whose execution time is several orders of magnitude longer than their scheduling delays, enabling sophisticated software schedulers with global coordination. At the microsecond timescale, however, the service time of tasks becomes comparable to or even shorter than a network RTT, making scheduling delay a non-negligible portion in the overall latency.

\rev{For a task with 10\,\textmu{}s service time, a centralized software scheduler deployed on its forwarding path adds an extra intra-rack round trip plus queuing that can easily exceed one-third of the total response time~\cite{barroso2017attack,singhvi2025falcon}, pushing tail latency beyond SLO thresholds~\cite{dean2013tail}. Distributed software schedulers avoid the extra hop but sacrifice global visibility, leading to suboptimal placement under high loads; neither approach delivers near-optimal scheduling at this timescale without prohibitive latency or host CPU overhead.}

\subsection{Theories behind In-Network Scheduling}
Scheduling behavior is largely determined by how tasks are queued and dispatched across workers. The underlying queuing model governs the trade-off between tail latency, work-conserving efficiency, and implementation complexity. 

Join-bounded-shortest-queue (JBSQ)~\cite{kogias2019r2p2} formalized this trade-off by combining a centralized scheduler queue with local queues on workers, and limited each worker's queue depth by a bound parameter $n$. When $n=1$, the scheduling becomes fully centralized and pull-based, \textit{i.e.}, each worker only fetches tasks when it is idle, avoiding HoL blocking but sacrificing work-conserving efficiency. At the opposite extreme, $n=\infty$ yields a fully distributed, push-based scheme that maximizes work conservation but can incur severe HoL blocking as workers' local queues can grow unbounded. The power-of-$k$ policy~\cite{mitzenmacher2001power} provided a non-global alternative by sampling $k$ workers per task and selecting the one with the shortest queue occupancy. This reduces coordination overhead due to state collection, though at the cost of less precise load balancing. The simulations in~\cite{kogias2019r2p2,daglis2019rpcvalet} suggested that centralized queuing combined with shallow worker queues provides the best tail latency, particularly under high service time dispersion. More broadly, schemes exploiting global queue information consistently outperformed those that rely on partial states.

\subsection{In-Network Scheduling in Practice}

\subsubsection{On-Switch Scheduling} \label{sec:on-switch_scheduing_in_practice}

Programmable switches such as Intel Tofino~\cite{intel2025tofino} provide high throughput and predictable, low per-packet latency, making them attractive substrates for in-network scheduling. However, translating the ideal principles derived from theory---centralized queuing and globally-coordinated scheduling---into the constraints of real hardware remains challenging.

Building a centralized queue requires buffering tasks of variable sizes. To access task metadata carried in packets, the switch has to parse packet fields, but packet parsers usually only support limited look-up lengths and thus cannot extract arbitrary \rev{task payload}. Moreover, the capability of maintaining a queue in on-chip memory is restricted by the narrow width and fixed structure of register arrays. Global coordination is also subject to hardware constraints. Specifically, in Tofino's feed-forward pipeline, each register array resides in a single pipeline stage and thus can only be accessed at most once per packet during a pipeline pass. This constraint precludes any iterative aggregation or comparison within the array. Consequently, when the number of queues exceeds the number of pipeline stages, scanning all the queues to find the globally shortest one cannot be done in a single pipeline pass, \emph{i.e.}, the queue registers are distributed across stages.

Due to these limitations, existing in-network schedulers adopt various compromises. R2P2~\cite{kogias2019r2p2} approximated centralized queuing through packet recirculation, keeping tasks in the pipeline instead of explicitly buffering them. Although this avoids generating large queues, the scheme actually consumes more switch bandwidth, increases latency, and risks packet reordering. In practice, we also observe that such recirculation causes the implemented behavior to deviate noticeably from the theoretical performance of JBSQ. Draconis~\cite{udayashankar2024draconis} is optimized for small, fixed-size tasks that can be buffered directly on the switch. For larger tasks, workers still need extra time to fetch the associated data after dispatch, which adds to their end-to-end latency. RackSched~\cite{zhu2020racksched} adopted a two-level scheduling design, in which the in-network component does not utilize centralized queuing and applies a sub-optimal power-of-$k$ policy.

\begin{figure}[h]
    \centering
    \includegraphics[width=0.75\textwidth]{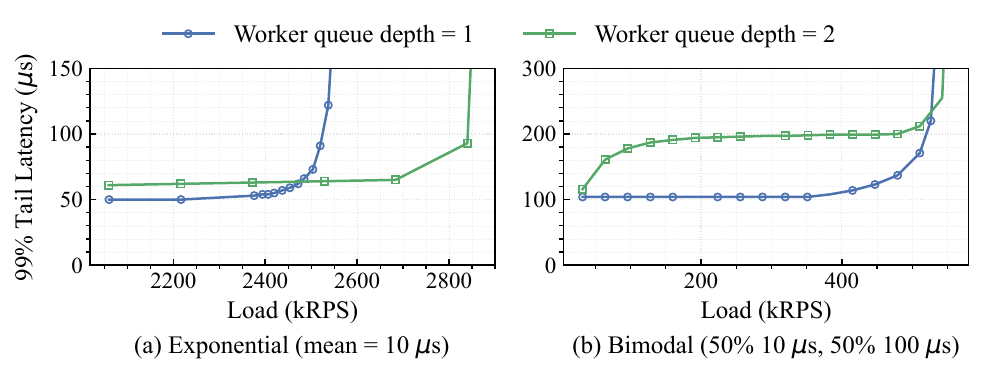}
    \caption{Tail latency \textit{versus} system load under two worker queue depths.}
    \label{fig:tail_lat_collapse}
\end{figure}

\subsubsection{Tail Latency Collapse} \label{sec:tail_latency_collapse}

A major discrepancy between theoretical queuing models and real-world systems lies in the cost of task retrieval. In theory, workers should be able to pull tasks from a centralized queue instantaneously, incurring no network delay or control overhead. However, in practice, each pull request has to traverse the network and compete for the processing resources on a switch or network interface card (NIC), and thus introduces non-negligible latency. Therefore, workers might stall while waiting for new tasks, making the system non-work-conserving~\cite{demoulin2021idling}. This inefficiency has a significant impact on stability. Specifically, as the load grows, accumulated pull delays reduce the system's effective service capacity, causing an earlier onset of \emph{tail-latency collapse}, \textit{i.e.}, a sharp increase in response time near saturation. As a consequence, shallow worker queues do not always improve tail latency: once the system enters the collapse region, tail latency escalates rapidly and soon outweighs the benefits of limiting queue depth.

Figures~\ref{fig:tail_lat_collapse}(a) and \ref{fig:tail_lat_collapse}(b) show the results when the service time of workloads follows exponential and bimodal distributions, respectively, where the queue depth of each worker is limited as 1 or 2 and the load is in kilo requests per second (kRPS). When the service time follows the exponential distribution, shallower queues collapse earlier than deeper ones, reflecting reduced work conservation and lower sustainable throughput. Under the bimodal distribution, this effect is less pronounced as the service time is more variable. In all, the results highlight that overly-short queues can magnify non-work-conserving behavior, making tail latency rise sharply as system approaching collapse.

\subsection{Classification Alleviates but Does Not Eliminate Dispersion}

A natural approach to mitigating service time dispersion is to classify tasks and route each type of tasks to a dedicated queue or worker pool~\cite{liao2025towards,demoulin2021idling,didona2019size}. This stratification narrows the service time distribution within each type and enables more informed scheduling decisions, reducing the chance that an exceptionally long task blocks many short ones. However, in real-world systems, execution time depends on numerous dynamic factors that have joint effects and thus can hardly be coarsely distinguished with type labels, including input data characteristics, cache locality, interference from co-located workloads, and variability in downstream storage or external services.

\begin{figure}[h]
    \centering
    \includegraphics[width=0.35\textwidth]{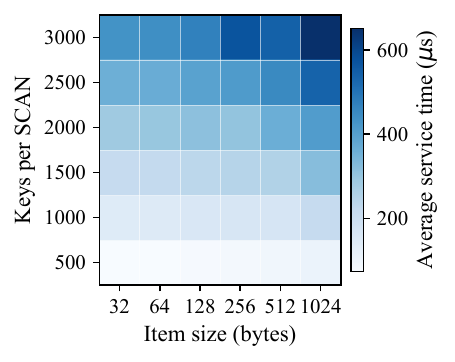}
    \caption{Average service time of \texttt{SCAN} in RocksDB as a function of item size and number of keys.}
    \label{fig:rocksdb_heatmap}
\end{figure}

For example, in RocksDB~\cite{rocksdb}, the service time of operations like \texttt{GET} and \texttt{SCAN} increases with the number of keys accessed, and it also correlates with the item size~\cite{didona2019size}. Our measurements in Figure~\ref{fig:rocksdb_heatmap} further confirm that these two factors jointly affect the average service time, even when we only consider a single operation type (\texttt{SCAN}). Hence, considerable residual dispersion can remain within each task type, amplifying the negative impact of HoL blocking. To this end, although classification is a useful primitive, it cannot fully eliminate workload dispersion, and thus practical schedulers have to address the residual variability among homogeneous tasks.

\section{Design of \sysname{}}

\subsection{System Overview}

\begin{figure}[h]
    \centering
    \includegraphics[width=0.65\textwidth]{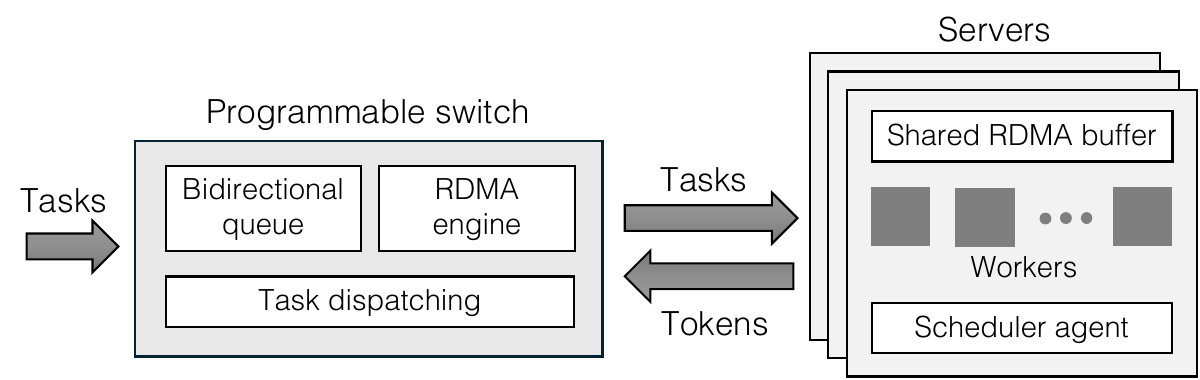}
    \caption{System architecture of \sysname{}.}
    \label{fig:system_architecture}
\end{figure}

The system of \sysname{} consists of two core components: a programmable switch and lightweight scheduler agents co-located with workers, as illustrated in Figure~\ref{fig:system_architecture}. Note that, we make \sysname{} focus on in-network scheduling and thus it is designed to be agnostic to worker roles, regardless of whether a worker serves as the final task handler or as a second-level scheduler~\cite{zhu2020racksched}.

The programmable switch serves as the ingress for worker traffic and performs per-task scheduling directly in the data plane. We design it to have the following modules: 1) a set of centralized bidirectional queues that are maintained for both tasks and tokens, enabling precise task dispatching~(\S\ref{sec:bidirectional_queue}), and 2) an RDMA engine that offloads large task payloads to workers' memory while retaining only compact metadata in the switch~(\S\ref{sec:rdma_engine}). The in-network queues and their corresponding workers are distributed to slices, such that tasks can be mapped to the slices according to their types, thereby reducing workload dispersion within each slice~(\S\ref{sec:scheduling_in_slices}). Each scheduler agent runs as a lightweight runtime module on a server, managing one or more workers. It monitors the runtime status of its workers, and adjusts their participation in scheduling accordingly, by tuning their token quotas and updating worker-to-slice mappings~(\S\ref{sec:adaptive_scheduling}).

In addition, we also include a controller for \sysname{}, which is not shown in Figure~\ref{fig:system_architecture}, to coordinate worker registration and program the switch with necessary metadata, such as the routing rules for task dispatching and RDMA configuration parameters that enable valid one-sided RDMA \texttt{WRITE}.

\subsection{On-Switch Scheduling} \label{sec:on-switch_scheduling}

\subsubsection{Bidirectional Queue} \label{sec:bidirectional_queue}

\begin{figure}[h]
    \centering
    \includegraphics[width=0.75\textwidth]{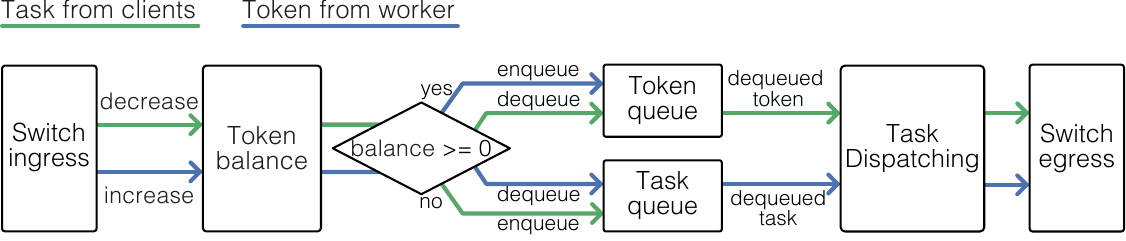}
    \caption{In-network scheduling workflow centered on the bidirectional queue.}
    \label{fig:bidirectional_queue}
\end{figure}

\sysname{} maintains in-switch queues for both client-submitted tasks and worker-issued tokens. When a task and a token cannot be immediately matched, the unmatched entry is buffered in the switch. At any time, only one side of the bidirectional queue, \emph{i.e.}, either the task queue or the token queue, can be non-empty. Figure~\ref{fig:bidirectional_queue} illustrates the overall workflow.

\textbf{Approximating JBSQ.}
As discussed in \S\ref{sec:on-switch_scheduing_in_practice}, the feed-forward architecture of programmable switches makes it difficult to efficiently enforce bounded local queues and to compute the global minimum queue length required for exact JBSQ. \sysname{} instead approximates JBSQ through a token-driven dispatching mechanism, which enforces bounded local queues as: each worker can hold at most $n$ outstanding tokens, and more lightly-loaded workers tend to retain more tokens and thus are more likely to receive new tasks. This mechanism coincides with JBSQ in two cases: 1) \rev{if task service time is constant}, it performs round-robin dispatching among the least-loaded workers, and 2) when each worker's local queue depth is limited to one, it becomes identical to JBSQ. In practice, \sysname{} further approaches JBSQ behavior by partitioning resources into slices to reduce workload dispersion~(\S\ref{sec:scheduling_in_slices}), and when dispersion is unavoidable, constraining the local queue depth to one again recovers the equivalence to JBSQ~(\S\ref{sec:adaptive_scheduling}).

\textbf{Efficient queue operations.}
To compactly encode the queue state, \sysname{} maintains a token-balance counter, where a positive value denotes the number of outstanding tokens, and a negative value (in absolute terms) is for the buffered tasks. This representation fits well with a feed-forward switch pipeline, as enqueue/dequeue operations are reduced to simple updates on the bidirectional queue. Hence, \sysname{} avoids costly check-and-rollback cycles with packet recirculation~\cite{udayashankar2024draconis}.

\textbf{Elimination of worker polling.}
By queuing tokens in the switch, \sysname{} removes the need of workers actively polling for tasks. Each worker issues a token to signal its readiness for a new task. The token is either immediately consumed by an outstanding task, or cached in the switch so that the next arriving task can be dispatched to the worker instantly. This push-based design maximizes responsiveness without incurring any additional bandwidth cost due to workers' polling.

\subsubsection{RDMA Engine} \label{sec:rdma_engine}

\begin{figure}[h]
    \centering
    \includegraphics[width=0.52\textwidth]{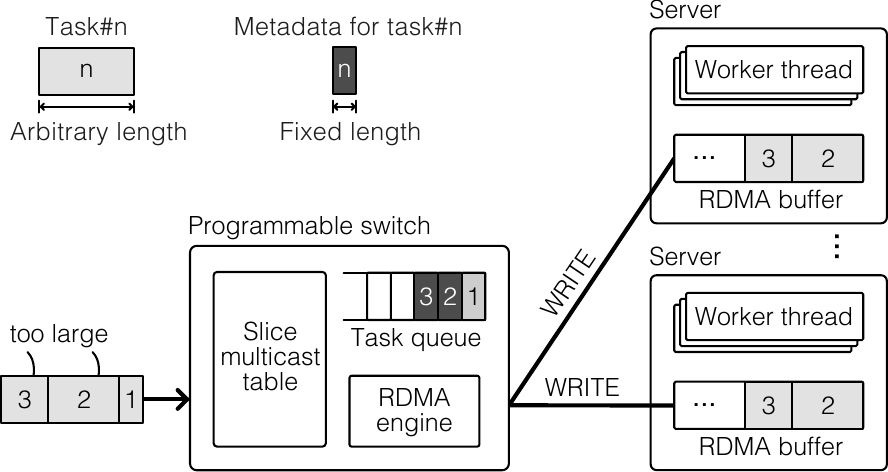}
    \caption{RDMA engine that pre-writes large tasks to worker memory via one-sided RDMA \texttt{WRITE}.}
    \label{fig:rdma_engine}
\end{figure}

Each task-queue entry in \sysname{} has a fixed length (\emph{e.g.}, 8\,bytes in our implementation) to ensure feasible implementation on programmable switches. However, this makes it difficult to accommodate the tasks whose sizes are larger than the length. Meanwhile, \sysname{} needs to make sure that all tasks participate in the same scheduling process without compromise. Therefore, we design a RDMA engine in \sysname{} to indirectly cache oversized tasks while preserving the abstraction of a centralized queue. As exemplified in Figure~\ref{fig:rdma_engine}, the tasks that cannot be directly buffered in the switch will be pre-written to the RDMA buffers in workers' memory via one-sided RDMA \texttt{WRITE} operations. Then, for each of these oversized tasks, the switch only needs to store in its task queue a fixed-size metadata descriptor, which records the remote memory location, payload size, and other per-task metadata. When the task is dispatched, the designated worker retrieves the full task content from its local RDMA buffer based on this metadata reference.

\textbf{\texttt{WRITE} multicasting.}
Supporting large tasks within the same scheduling pipeline means that the handling worker of a task is not known at the time of pre-writing. To address this, \sysname{} identifies all potential workers within the same scheduling slice~(\S\ref{sec:scheduling_in_slices}) and pre-writes payloads of oversized tasks to the RDMA buffer of each candidate worker via one-sided RDMA \texttt{WRITE} as follows. During initialization, each scheduler agent exposes a shared RDMA buffer and queue pair (QP) to its co-located workers. The RDMA buffer is organized as a ring, with the write offset maintained by the switch to ensure consistent wrap-around placement. Note that, this introduces no data races or synchronization overhead, as each worker accesses its buffer only after a task being dispatched to it. The switch uses its built-in multicast capability to perform RDMA \texttt{WRITE} multicasting.

\textbf{Constructing valid RDMA \texttt{WRITE} packets\footnote{\sysname{} targets RDMA over Converged Ethernet version~2 (RoCEv2)~\cite{wikipedia2025roce}.}.}
To construct valid RDMA \texttt{WRITE} packets from scratch, \sysname{} has to ensure the correctness of all the routing and RDMA state fields, including the destination MAC address, source and destination IP addresses, queue pair numbers (QPNs), and remote memory keys (rkeys). The packet sequence number (PSN) is also maintained continuously to preserve reliable connection semantics.
\rev{Tofino's packet replication engine creates one copy per multicast group member between the ingress and egress pipelines, and the per-destination RDMA fields are then rewritten by match-action tables keyed on the egress port, so each replica carries the correct per-worker RDMA state.}
The static connection state (\emph{e.g.}, routing information, QPNs, and rkeys) is stored in match-action tables installed by the controller during worker registration, while PSNs are dynamically tracked in a register array. We disable integrity checks (iCRC), as recomputing them in the switch pipeline is infeasible~\cite{ma2025rdmax}.

\subsubsection{Scheduling in Slices} \label{sec:scheduling_in_slices}

To mitigate workload dispersion, \sysname{} partitions both tasks and workers into slices. Both the bidirectional queue and RDMA \texttt{WRITE} multicasting operate at the granularity of slices. A data-plane table maps tasks to slice identifiers according to the tasks' types, and it is maintained and updated in runtime by the \sysname{} controller. During enqueue/dequeue operations, tasks and tokens consult this table to determine their slice. This design is based on the observation that tasks of the same type in an application tend to exhibit similar service time~\cite{liao2025towards}. For example, in RocksDB, \texttt{GET} operations typically access only a small amount of data (\emph{e.g.}, fetching $10$ keys costs about $12$ µs), whereas \texttt{SCAN} requests traverse much larger ranges (\emph{e.g.}, scanning $5,000$ keys costs about $650$ µs)~\cite{liao2025towards}. Hence, slicing groups similar tasks together such that scheduling decisions can be made among workers handling relative homogeneous workloads, thereby reducing the probability of HoL blocking. Note that, slicing is intended to alleviate (rather than eliminate) workload dispersion, as in certain cases, dispersion might still remain even after slicing~\cite{didona2019size}.

\rev{Note that the idea of type-based task separation has been explored before~\cite{liao2025towards,demoulin2021idling,didona2019size}; \sysname{} adopts it as an auxiliary optimization to reduce workload dispersion, while the key difference is that \sysname{} retains centralized bounded queuing within each slice, which bounds tail latency effectively.} 

\begin{algorithm}[t]
    \caption{Adaptive Scheduling}
    \label{alg:adaptive_scheduling}
    \begin{algorithmic}[1]
    \For{each sampling interval $\Delta t$} \label{line:loop-sampling-begin}
        \For{each slice $s$}
            \For{each worker $w\in\mathcal{W}_s$}
                \State sample $L_w$, $L_{sw}$, $T_w$, and $Q_{sw,w}$ for worker $w$;
                \State update $\mathcal{S}_w$, $\mathcal{R}_w$, $\mathcal{Q}_w$ with $\frac{L_w + L_{sw} + T_w}{T_w}$, $\frac{L_w}{L_w + L_{sw}}$, and $Q_{sw,w}$;  \label{line:loop-sampling-end}
            \EndFor
        \EndFor
    \EndFor

    \For{each control interval of $K$ sampling intervals} \label{line:loop-control}
        \For{each slice $s$} \label{line:loop-slices}
            \State $\text{label}(s)= \varnothing$, $O_s = 0$, $I_s = 0$; \label{line:init-s}
            \For{each worker $w\in\mathcal{W}_s$} \label{line:loop-workers}
                \State compute $99\%$ tail slowdown $S_w^{p99}$ from samples in $\mathcal{S}_w$; \label{line:slowdown-def}
                \State compute average ratio $\bar R_q$ from samples in $\mathcal{R}_w$; \label{line:rq-def}
                \State compute $99\%$ tail switch-side queue length $Q_{sw,w}^{p99}$ from samples in $\mathcal{Q}_w$; \label{line:qsw-def}
                \If{$S_w^{p99} > S_{\text{th}}$} \label{line:slowdown-high}
                    \If{($n_w>1$) \textbf{and} ($\bar R_q>R_{\text{th}}$) \textbf{and} ($Q_{sw,w}^{p99} = 0$)} \label{line:dec-cond}
                        \State $n_w = n_w-1$; \label{line:dec-n}
                    \ElsIf{($n_w<n_{\text{max}}$) \textbf{and} ($\bar R_q<R_{\text{th}}$)} \label{line:inc-cond}
                        \State $n_w = n_w+1$; \label{line:inc-n}
                    \Else
                        \State $O_s = O_s+1$; \label{line:Os-inc}
                    \EndIf
                \Else
                    \State measure idleness ratio of worker $w$ over the control interval as $I_w$;  \label{line:idle-ratio}
                    \State $I_s = I_s + I_w$; \label{line:Is-acc}
                \EndIf
                \State reset sample buffers $\mathcal{S}_w$, $\mathcal{R}_w$, $\mathcal{Q}_w$;
            \EndFor
            \If{($O_s>0$) \textbf{and} ($I_s>1$)} \label{line:both-ou}
                \State $N_s = \sum\limits_{w\in\mathcal{W}_s} n_w$, $\bar n = \frac{N_s}{|\mathcal{W}_s|}$; \label{line:avg-n}
                \For{each worker $w\in\mathcal{W}_s$} 
                    \State $n_w =\bar n$; 
                \EndFor \label{line:equalize-n}
            \ElsIf{$O_s>0$}   
                \State $\text{label}(s) = \textsc{Overloaded}$; \label{line:label-over}
            \ElsIf{$I_s>1$}   
                \State $\text{label}(s) = \textsc{Underloaded}$; \label{line:label-under}
            \EndIf
        \EndFor

        \State $\mathcal{S}_o = \{s\mid \text{label}(s)=\textsc{Overloaded}\}$, 
               $\mathcal{S}_u = \{s\mid \text{label}(s)=\textsc{Underloaded}\}$; \label{line:So-Su-def}
        \While{($\mathcal{S}_o\neq\emptyset$) \textbf{and} ($\mathcal{S}_u\neq\emptyset$)} \label{line:while-rebal}
            \State $s_o = \arg\max\limits_{s\in\mathcal{S}_o} (O_s)$, $s_u = \arg\max\limits_{s\in\mathcal{S}_u} (I_s)$; \label{line:select-slices}
            \State $w^\star = \arg\min\limits_{w\in\mathcal{W}_{s_u}} (n_w)$; \label{line:select-worker}
            \State reassign $w^\star$ to $s_o$, $s(w^\star)= s_o$, $n_{w^\star}= 1$; \label{line:reassign-worker}
            \State $\mathcal{S}_o = \mathcal{S}_o \setminus \{s_o\}$, $\mathcal{S}_u = \mathcal{S}_u \setminus \{s_u\}$; \label{line:update-Sets}
        \EndWhile
    \EndFor
    \end{algorithmic}
\end{algorithm}

\subsection{Adaptive Scheduling} \label{sec:adaptive_scheduling}

\sysname{} introduces an adaptive scheduling strategy to tune worker behaviors dynamically in runtime, \textit{i.e.}, continuously adjusting each worker's token quota and worker-to-slice mapping in response to changes in workload characteristics and performance variations caused by interference~\cite{fried2020caladan}. Specifically, to avoid tail latency collapse~(\S\ref{sec:tail_latency_collapse}), we make sure that the adaptive scheduling can distinguish the rise in tail latency stemming from overload-induced collapse from that due to HoL blocking. \sysname{} infers the root-cause by examining the ratio of worker-side queuing latency to total queuing latency, \textit{i.e.}, HoL blocking driven by deep worker queues can usually be identified by seeing a disproportionate increase in worker-side queuing delay.

To implement this idea, we make each scheduler agent continuously collect switch-side and worker-side queuing latencies ($L_{sw}$ and $L_w$, respectively). To avoid requiring clock synchronization between the switch and NICs on servers, the latencies are measured independently in their respective clock domains and then aggregated as durations to obtain the total queuing delay. In addition, it records processing times ($T_w$) and switch-side queue occupancy ($Q_{sw,w}$) observed by the worker. All switch-side metrics are embedded into packets via in-band network telemetry (INT)~\cite{kreonet2025int}. To support adaptive worker-to-slice allocation, each worker also reports its idleness. For threads using a poll-mode driver (PMD)~\cite{dpdk2025pmd}, the ratio of polling iterations that return zero packet to total iterations provides an effective measure of idleness~\cite{intel2023core}.

\textit{Algorithm}~\ref{alg:adaptive_scheduling} describes the procedure of the adaptive scheduling, which is implemented by scheduler agents running alongside workers and invoked at a high frequency. The agent continuously runs a fine-grained sampling loop (\textit{Lines}~\ref{line:loop-sampling-begin}--\ref{line:loop-sampling-end}) that, every small sampling interval $\Delta t$ (\emph{e.g.}, 10\,$\mu$s), collects measurements $\{L_w,L_{sw},T_w,Q_{sw,w}\}$ from all slices and workers and appends the derived statistics to per-worker sample buffers. Every control interval consisting of $K$ such sampling steps (\textit{Line}~\ref{line:loop-control}). At the per-worker level (\textit{Lines}~\ref{line:loop-workers}--\ref{line:Is-acc}), the agent computes from the accumulated samples the $99\%$ tail slowdown $S_w^{p99}$ (\textit{Line}~\ref{line:slowdown-def}), the average ratio $\bar R_q$ of worker-side latency to total queuing latency (\textit{Line}~\ref{line:rq-def}), and the $99\%$ tail switch-side queue length $Q_{sw,w}^{p99}$ (\textit{Line}~\ref{line:qsw-def}).

When the slowdown $S_w^{p99}$ exceeds the preset threshold $S_{\text{th}}$ (\textit{Line}~\ref{line:slowdown-high}), the agent first determines where the queuing is dominated. If it is dominated by the worker queue ($\bar R_q > R_{\text{th}}$) while the switch queue remains empty at the $99\%$ tail level ($Q_{sw,w}^{p99}=0$) and the worker still has more than one outstanding token ($n_w>1$), the agent conservatively decreases the worker's token quota $n_w$ by one (\textit{Line}~\ref{line:dec-n}) to mitigate HoL blocking. Symmetrically, if queuing is instead dominated by the switch queue (\emph{i.e.}, $\bar R_q < R_{\text{th}}$) and the worker has not reached its maximum quota ($n_w<n_{\text{max}}$), the agent cautiously increases $n_w$ by one (\textit{Lines}~\ref{line:inc-cond}--\ref{line:inc-n}) to help drain the switch queue. Only when neither of these cases applies does the agent interpret the event as persistent overload and increment the slice's overload counter $O_s$ (\textit{Line}~\ref{line:Os-inc}).

When the slowdown stays below the threshold ($S_w^{p99}\leq S_{\text{th}}$), the agent aggregates the idleness of workers in a slice by summing up their idleness ratios $I_w$ over the entire control interval (\textit{Lines}~\ref{line:idle-ratio}--\ref{line:Is-acc}). After checking all the workers in a slice, the agent uses $O_s$ and the total idleness $I_s$ to infer the slice state (\textit{Lines}~\ref{line:both-ou}--\ref{line:label-under}). If both overload and underutilization signals are present, it equalizes local depths by averaging $n_w$ within the slice (\textit{Lines}~\ref{line:avg-n}--\ref{line:equalize-n}), smoothing intra-slice imbalance without changing the slice's worker set. Otherwise, it labels the slice as overloaded or underloaded for potential cross-slice rebalancing; a slice is considered underloaded only when the total idleness exceeds one worker-equivalent ($I_s>1$), so that reassigning a worker is relatively safe in terms of not overloading the slice. The global rebalancing routine proceeds greedily (\textit{Lines}~\ref{line:So-Su-def}--\ref{line:update-Sets}): in each iteration, it pairs the most overloaded slice and the slice with the largest total idleness (\textit{Line}~\ref{line:select-slices}), migrates the least-provisioned worker from the latter to the former (\textit{Lines}~\ref{line:select-worker}--\ref{line:reassign-worker}), and updates the candidate sets accordingly (\textit{Line}~\ref{line:update-Sets}). The migrated worker's token quota is initialized to $n_{w^\star}=1$, after which subsequent epochs adjust its depth as needed.

The scheduler agent interacts with workers via shared memory: workers expose their local statistics in a shared control block, and the agent updates control knobs. Per-worker control state is laid out on separate cache lines to avoid inter-worker false sharing~\cite{wikipedia2025false}. We use a sampling interval of $\Delta t=10$\,$\mu$s and aggregate $K=1,000$ samples per worker before each control decision. The slowdown threshold for triggering token-quota adjustments is set to $S_{\text{th}}=10$, with the token quota capped at $n_{\text{max}}=8$. We will analyze the effect of ratio threshold $R_{\text{th}}$, which distinguishes worker-queue-dominated slowdown from switch-queue-dominated slowdown in evaluations (\S\ref{sec:dynamic_evaluation}).

\rev{
\textbf{Relation to backpressure.}
Unlike classical backpressure, which reacts monotonically to queue buildup, \textit{Algorithm}~\ref{alg:adaptive_scheduling} first determines the root cause of tail-latency degradation via the ratio $\bar R_q$ and then takes proper actions to address either HoL blocking or tail-latency collapse~(\S\ref{sec:tail_latency_collapse}).
}

\section{Implementation} \label{sec:implementation}

We prototype \sysname{} using an Intel Tofino switch~\cite{intel2025tofino} and commodity servers equipped with NVIDIA ConnectX-5 100\,GbE NICs. On the switch, our implementation consumes 11 out of the 12 pipeline stages and uses 24.1\% of the available SRAM without relying on any TCAM resources. On the server side, we use the data plane development kit (DPDK)~\cite{dpdk2025} to implement high-performance packet I/O, including flow steering for dispatching tasks to their designated workers.

\rev{Tofino's per-stage memory is constrained, so we maximize queue depth by striping each queue entry across pipeline stages, storing only one byte per stage. Under this layout, the in-switch payload width is bounded by the number of stages dedicated to it: we allocate eight of them, capping in-switch task data at 8\,bytes per entry alongside the required metadata (\emph{e.g.}, the client identifier). Note that, this 8-byte cap is an implementation limitation due to our pipeline stage budget rather than a design choice driven by application requirements. Under this design, the implementation supports up to 128\,K queued task entries.}

Although we optimize queue depth aggressively, queue overflow under heavy load with significant workload dispersion cannot be eliminated. \rev{As a safeguard, we cap the total outstanding requests at $Q_{\max} = Q_{sw} + \sum\limits_w n_w$, where $Q_{sw}$ is the switch queue capacity and $n_w$ is the token quota of worker~$w$. Scheduler agents distribute each client's share of $Q_{\max}$ via the control path, and clients enforce admission control locally and apply backpressure to the application layer, preventing silent drops and switch-side crashes.} Note that, the adaptive scheduling is not an overload-control mechanism because it only regulates worker-side queues without limiting outstanding tasks globally.

\section{Evaluations}

We evaluate \sysname{} with a series of experiments designed to answer the following questions:

\begin{itemize}
      \item How does \sysname{} compare to the SOTAs of in-network schedulers in terms of tail latency and service time slowdown under static workloads? (\S\ref{sec:static_evaluation})
      \item How does \sysname{}'s adaptive scheduling perform under dynamic workloads? (\S\ref{sec:dynamic_evaluation})
      \item \rev{What extra packet rate and buffer footprint does \sysname{}'s RDMA pre-writing incur? (\S\ref{sec:rdma_overhead})}
\end{itemize}

\subsection{Evaluation Setup}

Our testbed consists of three servers directly connected to a programmable switch based on Intel Tofino ASICs. Each server is equipped with Intel Xeon Silver 4316 CPUs and a 100-Gbps NVIDIA ConnectX-5 NIC configured in the Ethernet mode. The NIC and all CPU cores used in our experiments reside on the same non-uniform memory access (NUMA) node to avoid cross-NUMA traffic. All the servers run Linux kernel 5.15.0 and use DPDK 22.11.4 to accelerate packet I/O.

Two servers act as workers and one operates as the client. Both the client and workers are implemented as multi-threaded applications. Client threads generate workloads and collect throughput and latency statistics. Worker threads follow a symmetric thread model, \emph{i.e.}, each maintains a first-in-first-out (FIFO) task queue and executes tasks in a run-to-completion fashion. By default, we allocate 16 CPU cores to the client and 32 CPU cores to workers totally (across the two servers).

\subsection{Evaluations with Static Workloads} \label{sec:static_evaluation}

We first evaluate \sysname{} under static workloads whose statistical properties remain constant, and vary only the offered load to obtain load-latency curves. Each data point is obtained by first running static workloads at the selected load for one second for warming up, followed by a measurement interval of 10 seconds. The benchmark systems are as follows:

\begin{itemize}
    \item \textbf{R2P2~\cite{kogias2019r2p2}:}  
    It implements the JBSQ($n$) strategy using packet recirculation, and relies on explicit queue-depth reports from workers to keep its on-switch cache updated. Specifically, each task packet traverses the switch pipeline iteratively, sampling a few workers in each pass to determine whether a worker meets the target queue depth. Tasks begin by looking for workers with a queue depth of one. If none can be found after checking all the workers, the target queue depth might be increased by one\footnote{We use either JBSQ(1) or JBSQ(2), depending on the specific workloads under evaluation.}.

    \item \textbf{RackSched~\cite{zhu2020racksched}:}  
    It performs on-switch scheduling using a power-of-$k$ approach\footnote{We set $k=2$ in the evaluations.}. Upon the arrival of a task, the switch randomly selects a subset of workers and assigns the task to the worker with the shortest reported queue depth, without imposing an upper-bound on queue length. RackSched either depends on explicit worker reports or incurs an extra recirculation pass after each worker selection, to refresh the on-switch queue-depth cache.

    \item \textbf{Draconis~\cite{udayashankar2024draconis}:}  
    It buffers tasks directly in the switch and adopts a pull-based model where workers fetch tasks proactively, to approximate JBSQ(1). When workers are idle and need new tasks, they have to continuously issue pull requests to retrieve the tasks.

    \item \textbf{Pallas~\cite{liao2025towards}:}  
    It isolates tasks by type and dispatches tasks in each type to a dedicated subset of workers. For queue management, it uses a simple weighted round-robin policy to select workers and directly pushes tasks without limiting worker queue depth.
\end{itemize}

Our performance comparisons focus mainly on in-network scheduling, and thus all the experiments use the same symmetric worker-thread model. Note that, some baseline systems encompass more than just an in-network scheduler, \textit{e.g.}, RackSched also includes an intra-server scheduler. To keep the discussions focused, we use the suffix ``SW'' to refer specifically to their switch-side implementations throughout the rest of this section.

\subsubsection{Synthetic Workloads}

\begin{figure}[h]
    \centering
    \includegraphics[width=0.95\textwidth]{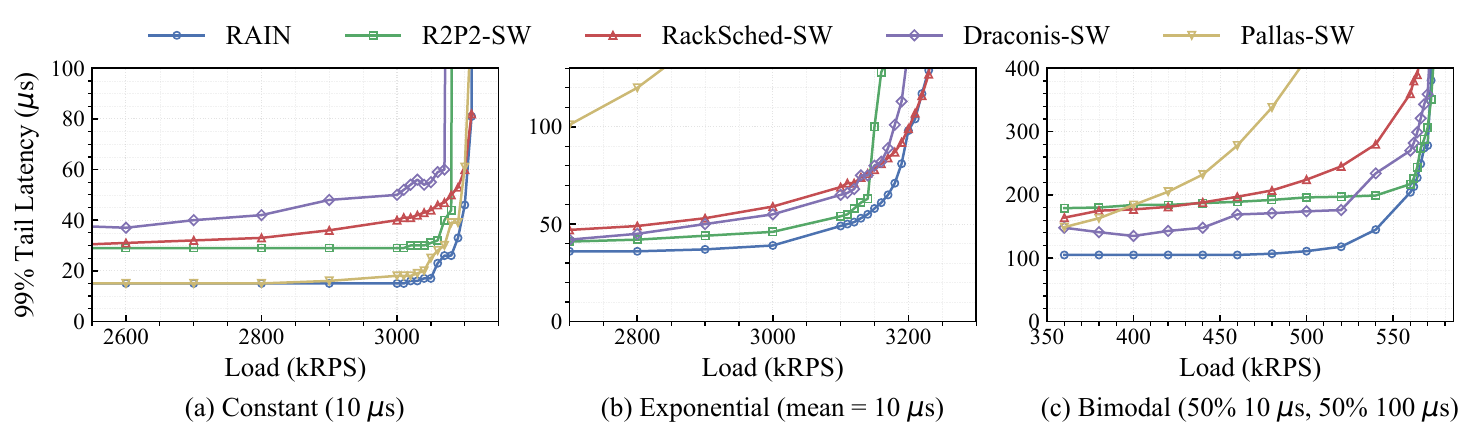}
    \caption{Tail latency \textit{versus} system load under static synthetic workloads.}
    \label{fig:synthetic_workloads}
\end{figure}

We first evaluate \sysname{} using synthetic workloads with constant, exponential, and bimodal service-time distributions, and the results are shown in Figures~\ref{fig:synthetic_workloads}(a)-\ref{fig:synthetic_workloads}(c), respectively. \rev{In this setting, we disable slice partitioning to isolate the contribution of \sysname{}'s bidirectional queuing mechanism.} We do not distinguish task types, \emph{i.e.}, any task can be dispatched to any worker, and the switch-side queue is shared by all the tasks. This corresponds to a scenario where slice partitioning by task types is not feasible, and thus neither \sysname{} nor Pallas can apply task-type-aware scheduling in this evaluation. The sizes of $80\%$ of the tasks are 8-byte, while the remaining can use a longer size. For tasks larger than 8\,bytes, \sysname{} transfers payloads via RDMA multicasting, whereas Draconis incurs an additional delay (equal to a RTT) to retrieve task data.

\textbf{Observation 1: \sysname{} achieves the best performance in tail-latency.}
In Figure~\ref{fig:synthetic_workloads}(a) (with constant service time), both \sysname{} and Pallas-SW achieve the shortest tail latency, when the load does not exceed 2,800\,kRPS. As workload dispersion is negligible in this case, Pallas-SW's round-robin and push-based policy does not induce noticeable HoL blocking, \textit{i.e.}, it effectively approximates the JBSQ(1) behavior provided by \sysname{} at relatively low loads. However, as the load further increases or workload dispersion grows, the disadvantages of Pallas-SW become evident, and HoL blocking causes its tail latency to rise much faster, making it perform the worst in Figures~\ref{fig:synthetic_workloads}(b) and \ref{fig:synthetic_workloads}(c).

Interestingly, although R2P2-SW and Draconis-SW also implement JBSQ(1), they both provide noticeably longer tail latency than \sysname{}. In Figure~\ref{fig:synthetic_workloads}(a), when the load does not exceed 2,800\,kRPS, \sysname{} provides $\sim$15\,µs tail latency, whereas that from R2P2-SW and Draconis-SW is roughly 30\,µs and 35\,µs, respectively. We attribute this gap to the overheads in the respective JBSQ(1) implementations of R2P2-SW and Draconis-SW. R2P2-SW heavily relies on recirculation, which can lead to task reordering, \textit{i.e.}, tasks undergoing recirculation might be overtaken by later arrivals that more quickly seize the opportunity to be scheduled to a worker, causing certain tasks to experience unexpectedly large numbers of recirculations and thus prolonging tail latency significantly. For Draconis-SW, workers pull tasks continuously, and even through we maximize the pull frequency without considering bandwidth overhead, the explicit pulling still diverges from the idea behavior of JBSQ(1). Moreover, Draconis-SW incurs extra fetch time for large tasks (workers need to retrieve task data after being dispatched metadata), further widening the tail latency gap relative to \sysname{}.

Push-based systems (RackSched-SW and Pallas-SW) fall behind pull-based designs (\sysname{} and Draconis-SW), under highly-dispersive bimodal workloads. RackSched-SW's power-of-$k$ policy and push-based dispatch perform well and can even surpass certain pull-based designs, only when both workload dispersion and load are low, attributing to its relatively low implementation overhead.

\textbf{Observation 2: \sysname{} fully utilizes system throughput.}
Pull-based schedulers normally offer better tail latency, but their non-work-conserving behavior, caused by fixed worker queue depths, can make them susceptible to tail latency collapse as load increases, which will significantly offset their advantage on tail latency. Specifically, both R2P2-SW and Draconis-SW use statically-configured worker queue depths, which can hardly adapt to sudden load peaks, making them the system bottleneck, even the systems are ready to deliver higher throughput with deeper queues.

In contrast, \sysname{}'s adaptive scheduling dynamically adjusts worker queue depths by monitoring signs of HoL blocking in runtime, thereby avoiding the inherent limitations of aforementioned pull-based designs. For example, for the test cases in Figure~\ref{fig:synthetic_workloads}, \sysname{} increases the worker queue depth to 8 to adapt to high loads in the scenario with constant service time, whereas when the service time is under the bimodal distribution, it expands the queue depth more conservatively to 2 for heavy loads. We will further analyze the behavior and impact of adaptive scheduling in \S\ref{sec:dynamic_evaluation}.

\begin{figure}[h]
    \centering
    \includegraphics[width=.98\textwidth]{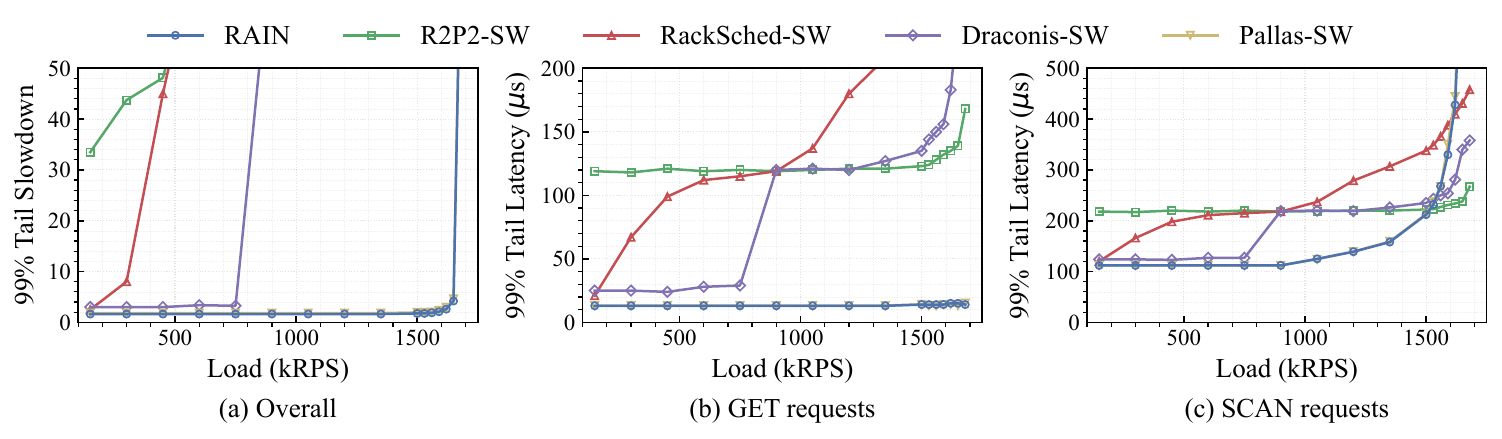}
    \caption{Overall $99\%$ tail slowdown and per-class $99\%$ tail latency (fixed number of keys per request).}
    \label{fig:rocksdb_const}
\end{figure}

\subsubsection{Workloads of Real-world Application}
We further evaluate \sysname{} using a real-world application, RocksDB~\cite{rocksdb}, a widely deployed key-value store. In this evaluation, we enable task classification, as RocksDB requests naturally fall in distinct categories and exhibit substantial differences in average service time. By default, the workloads consist of 90\% \texttt{GET} and 10\% \texttt{SCAN} requests, where each \texttt{GET} \rev{request packet explicitly carries 10 keys, which the worker iteratively serves via 10 point \texttt{Get} operations} with a key size of 64\,bytes and each \texttt{SCAN} \rev{request specifies a range of 500 keys (a starting key plus a count), which the worker serves via iterator traversal} with a key size of 1,024\,bytes, taking approximately 8\,$\mu$s and 107\,$\mu$s in our setup, respectively. In both \sysname{} and Pallas-SW, \texttt{GET} and \texttt{SCAN} requests use independent on-switch queues and dedicated workers (14 and 18 workers, respectively), while all the other systems use a common pool of workers. The results are shown in Figure~\ref{fig:rocksdb_const}, including the 99\% tail latency for \texttt{GET} and \texttt{SCAN} requests and the overall 99\% tail slowdown, defined as the completion time normalized by service time.

\textbf{Observation 3: Scheduling in slices delivers substantial tail-latency benefits.}
\rev{By enabling task classification on top of bidirectional queuing, this experiment isolates the additional benefit of slice-aware scheduling.}
Since both \sysname{} and Pallas-SW classify workloads to process them separately, they achieve significantly lower tail latency than those without. Specifically, they can keep the overall 99\% tail slowdown below 10 even when the load reaches 1,650\,kRPS, whereas the best-performing design without classification (\textit{i.e.}, Draconis-SW) can only maintain the slowdown below 10 when the load is below 900\,kRPS.

Although Draconis-SW does not perform classification, its JBSQ(1) policy allows it to maintain reasonable performance under moderate load. However, its non-work-conserving behavior, combined with the highly-dispersive workloads used in this evaluation, makes it prone to overload and tail-latency collapse. For R2P2-SW, we observe that the tail latency of its \texttt{GET} tasks approaches to the service time of \texttt{SCAN} tasks even under relatively light loads. This further supports our earlier observation that its recirculation-based JBSQ implementation introduces task reordering, which in turn causes the tail latency to deviate from its theoretical prediction. As expected, the tail latency from RackSched-SW, which relies on a push-based strategy and imposes no limit on worker queue depth, increases rapidly with load for both task types, causing a substantial rise in overall slowdown.

The downside of scheduling in slices is reduced peak throughput. Since \sysname{} and Pallas-SW divide workers across task types, their maximum achievable throughput is lower than that of systems with shared workers, which is reflected by the earlier overload point of \texttt{SCAN} requests shown in Figure~\ref{fig:rocksdb_const}(c). This reduction in peak throughput is an inherent cost of slice-based scheduling, but the benefit in overall 99\% tail slowdown remains substantial, justifying its necessity clearly.

\begin{figure}[h]
    \centering
    \includegraphics[width=.98\textwidth]{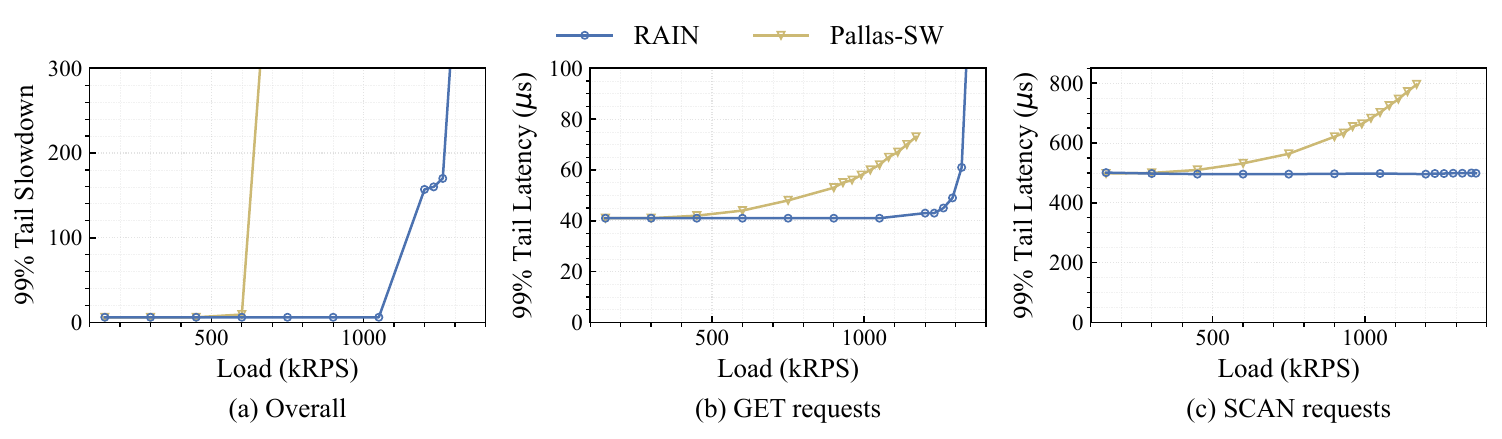}
    \caption{Overall 99\% slowdown and per-class 99\% tail latency (keys per request in exponential distribution).}
    \label{fig:rocksdb_exp}
\end{figure}

\textbf{Observation 4: \sysname{} effectively adapts to real-world workloads.}
In the previous evaluation, both the number of keys and item sizes were fixed for each task type, resulting in nearly constant per-class service time. This explains why the push-based Pallas-SW performed similarly to \sysname{}. To evaluate with a more realistic scenario, we draw the numbers of keys per \texttt{GET} and \texttt{SCAN} request from exponential distributions with means of 10 and 500 keys, respectively, and repeat the tests. 

The results are plotted in Figure~\ref{fig:rocksdb_exp}. Consistent with our observations in the tests with synthetic workloads, when Pallas-SW cannot rely on classification to control task dispersion, its push-based queuing strategy makes tail latency rise rapidly. In contrast, \sysname{} keeps regulating worker queue depth to avoid HoL blocking. Therefore, \sysname{} achieves 1.75$\times$ higher throughput than Pallas-SW before its 99\% tail slowdown exceeds 10, and at the load where Pallas-SW approaches overloaded, \sysname{} reduces the 99\% tail latency by 41.10\% and 37.69\% for \texttt{GET} and \texttt{SCAN} requests, respectively.

\begin{figure}[h]
    \centering
    \includegraphics[width=0.98\textwidth]{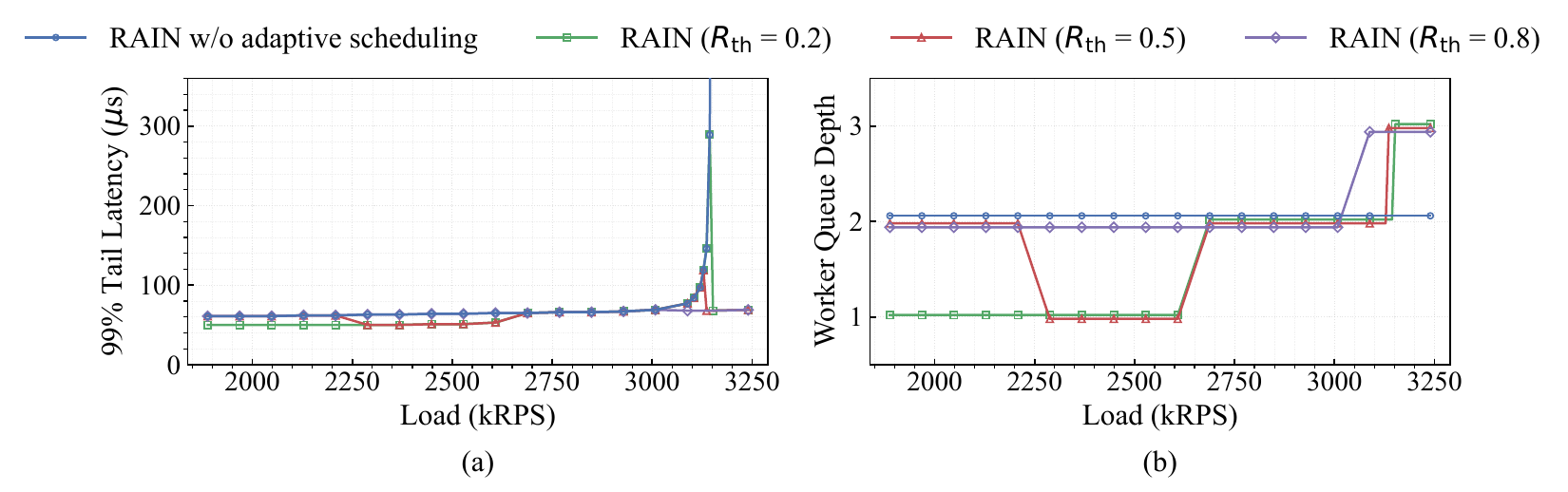}
    \caption{\rev{Impact of HoL-blocking threshold $R_{\text{th}}$ on \sysname{}'s adaptive scheduling\protect\footnotemark.}}
    \label{fig:parameter_tunning}
\end{figure}
\footnotetext{\rev{Queue depths in (b) are integer; curves offset vertically for clarity.}}

\subsection{Evaluations with Dynamic Workloads} \label{sec:dynamic_evaluation}

In this section, we evaluate \sysname{}'s adaptive scheduling under dynamic workloads. \rev{These experiments isolate adaptive scheduling's contribution beyond the static mechanisms above.}

\subsubsection{Parameter Tuning}

\sysname{}'s adaptive scheduler exposes two key parameters: the target 99\% tail slowdown $S_{\text{th}}$ and the threshold $R_{\text{th}}$. Since the former ($S_{\text{th}}$) is typically dictated by user requirements, we fix it as 10, while the latter ($R_{\text{th}}$) limits the fraction of queuing delay attributable to worker-side HoL blocking. Using the workloads with exponentially distributed service time (mean value at 10\,$\mu$s), we initialize the worker queue depth as 2 and gradually increase system load to test different values of $R_{\text{th}}$. The results are shown in Figure~\ref{fig:parameter_tunning}.

\textbf{Observation 5: $R_{\text{th}}$ governs the tradeoff between tail latency and system stability.}
A smaller $R_{\text{th}}$ lets \sysname{} tolerate less HoL blocking, making it more aggressive in reducing the worker queue depth $n$ to control HoL blocking, while a larger~$R_{\text{th}}$ yields more conservative adjustments. As shown in Figure~\ref{fig:parameter_tunning}, when $R_{\text{th}} = 0.2$, the system reduces $n$ from 2 to 1 at relatively low load, achieving the best tail latency when the load is below 2,300\,kRPS. However, this aggressive reduction makes the system more susceptible to overload and tail-latency collapse at higher loads. In contrast, when we set $R_{\text{th}} = 0.8$, the queue depth remains unchanged until the load reaches 3,000\,kRPS and is increased only when the system approaches overloaded. In summary, a smaller value of $R_{\text{th}}$ favors lower tail latency, while increasing its value makes the system emphasize more on stability. 

\begin{figure}[h]
    \centering
    \includegraphics[width=0.5\textwidth]{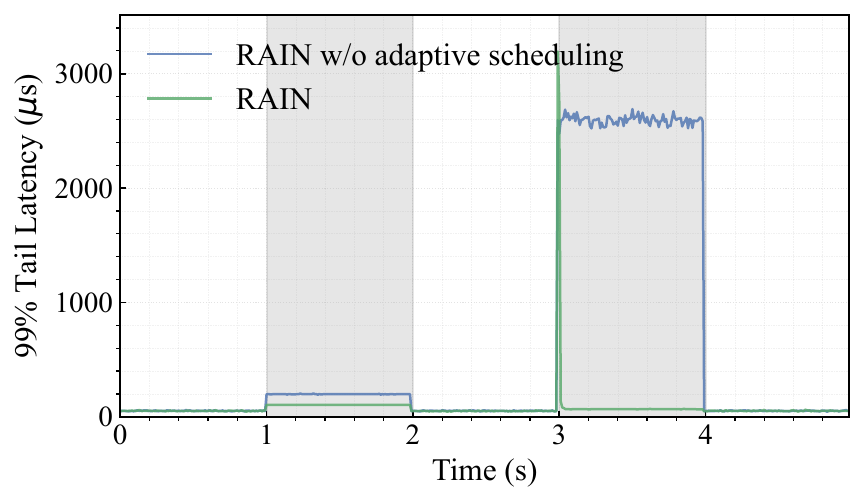}
    \caption{Real-time 99\% tail latency sampled every 10\,ms. At $t=1$\,s, the dispersion of input workloads increases for 1\,s before returning to normal, and then at $t=3$\,s, the system load increases for 1\,s and then falls back.}
    \label{fig:adaptive_scheduling}
\end{figure}

\subsubsection{Time-varying Workloads}
Finally, we consider a scenario with time-varying workloads, in which both the offered load and the dispersion of workloads can change dynamically. We set $R_{\text{th}}$ as 0.3 and monitor the tail latency to evaluate the effectiveness of \sysname{}'s adaptive scheduling. Initially, the workers are evenly divided into two slices, and then their allocations are adjusted by the adaptive scheduling. The worker queue depth is initialized as 2. We inject workloads into one of the slices, and Figure~\ref{fig:adaptive_scheduling} shows the results from the working slice.

\textbf{Observation 6: \sysname{}'s adaptive scheduling maintains low tail latency for time-varying workloads.}
We begin with workloads whose service time is exponentially distributed (mean value at 10\,$\mu$s) at a very light system load of 160\,kRPS. As the system is far from saturation and the slowdown remains small, adaptive scheduling does not intervene. At $t = 1$\,s, we increase the dispersion of workloads by switching to a bimodal distribution with 50\% of tasks having a service time of 10\,$\mu$s and the service time of remaining 50\% at 100\,$\mu$s. Hence, although the system is still underutilized, the degree of HoL blocking rises significantly, and in response, the adaptive scheduling reduces the worker queue depth from 2 to 1, thereby improving tail latency.

At $t = 3$ s, we increase the offered load to 1,600\,kRPS, pushing the system to overloaded and causing tail latency to spike. Over the next five control intervals (each lasting 10 ms), \sysname{}'s adaptive scheduler performs the following adjustments in sequence: it 1) restores the queue depth from 1 to 2, 2) further increases the queue depth to 3, 3) moves one worker from the idle slice to the working one, when observing that the tail slowdown remains above the target, 4) borrows the second worker from the idle slice, after which the tail latency returns to within the bound of target slowdown, and 5) reduces the queue depth to 2, when detecting that, with two additional workers, the switch queue stays at zero while the worker-side HoL blocking remains high. These adjustments enable \sysname{} to respond promptly to workload dynamics and maintain stable tail latency.

\rev{
\subsection{RDMA Pre-writing Overhead Analysis} \label{sec:rdma_overhead}

\sysname{} pre-writes a task's payload via RDMA multicasting only when the payload exceeds the on-switch cache capacity (\emph{e.g.}, larger than 8\,bytes) \emph{and} no free worker token is available upon arrival; otherwise, the task follows the normal forwarding path. The aggregate RDMA write rate is thus $R_{\text{RDMA}} = T \cdot \alpha \cdot \beta \cdot m$, where $T$ is task throughput, $\alpha$ the fraction of tasks buffered in the on-switch queue, $\beta$ the fraction whose payload cannot be cached, and $m$ the number of candidate workers per multicast. Since $\beta$ and $m$ are fixed by application and deployment, $\alpha$ is the only runtime knob. To stress-test it, we replay the three synthetic workloads of \S\ref{sec:static_evaluation} with all tasks enlarged to 16\,bytes ($\beta = 1$) and $m = 2$, with adaptive scheduling (\S\ref{sec:adaptive_scheduling}) enabled and queue depth ranging from 1 to 4. Figure~\ref{fig:queue_ratio} reports the measured $\alpha$.

\textbf{Observation 7: \sysname{}'s RDMA pre-writing incurs bounded overhead in both write rate and memory.}
At low-to-moderate loads, $\alpha$ stays negligible across all three workloads, as arriving tasks almost always find a free token and bypass on-switch queuing. Near saturation $\alpha$ rises, but adaptive scheduling counteracts this by enlarging the worker queue depth to replenish free tokens. Non-trivial overhead thus arises only near saturation with a large fraction of large-payload tasks, which remains tolerable since microsecond-scale applications are typically IOPS-bounded rather than bandwidth-bounded; if tighter bounds are required, operators can cap slice size or restrict multicasting to a subset of servers per slice. The per-server memory overhead is likewise bounded, since each worker only needs to reserve a small pre-write buffer proportional to its adaptive queue depth (at most 8 in our evaluations) and the largest task payload, which is negligible compared to the memory already provisioned on modern servers.

\begin{figure}[h]
    \centering
    \includegraphics[width=0.98\textwidth]{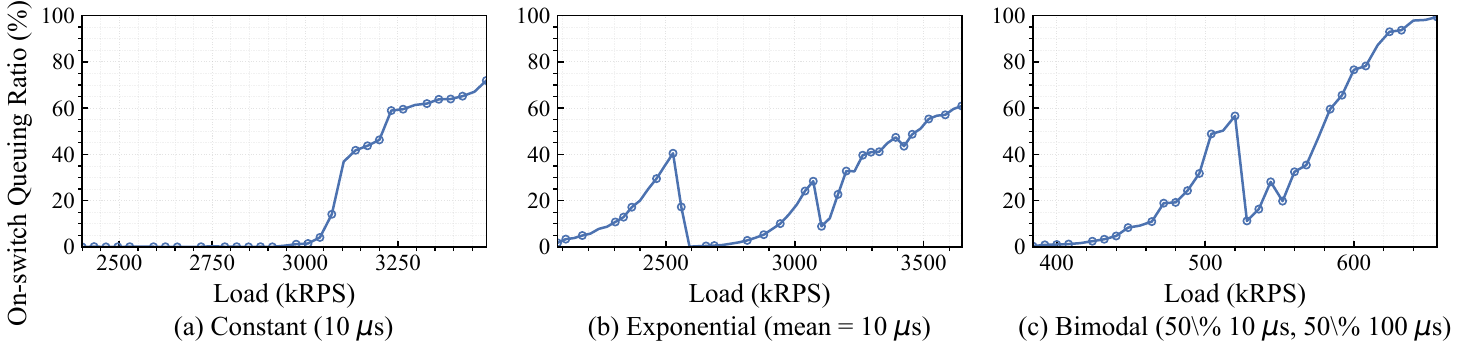}
    \caption{\rev{On-switch queuing ratio \textit{versus} system load under static synthetic workloads.}}
    \label{fig:queue_ratio}
\end{figure}

}
\rev{
\section{Discussion}
}

\rev{
\subsection{Failure Handling and Correctness of the RDMA Engine}
\label{sec:rdma_failure_correctness}

\sysname{} targets a single-rack deployment where the ToR switch also play the role of in-network scheduler, forming a controlled RDMA domain (\emph{e.g.}, RoCEv2); transient packet loss, NIC/QP errors, and worker or switch restarts can still occur, and our goal is to bound their impact rather than mask all failures in the data plane. For correctness, the outstanding tokens per node are capped by the RDMA ring-buffer size to prevent overrun, and each RDMA-written payload carries a copy of the request metadata so a worker executes a task only when the switch-delivered descriptor matches the buffer slot, preventing consuming stale or mismatched payloads. For recovery, a worker detecting a QP error stops issuing tokens and notifies the scheduler agent, which deregisters the worker and resets its RDMA state (QPN, PSN, rkey, buffer offset, tokens); once the QP is re-established, the agent registers the fresh state and the worker rejoins its slice. Since state is tracked per worker, a failure only affects that worker's outstanding tokens while other workers and slices are not interrupted; orphaned tasks are recovered via end-host timeout and resubmission, and unused RDMA replicas on non-selected candidates are reclaimed naturally as the ring buffer advances.

}

\rev{
\subsection{RDMA MTU and Multi-Packet Requests}

By configuring the host networking MTU to be no larger than the RDMA MTU (typically 4\,KB), each RDMA WRITE fits within a single RDMA packet, and the switch never needs to split one WRITE into multiple operations. When a logical request exceeds one MTU, the end-host network stack segments it into multiple packets before transmission. To maintain request affinity, \sysname{} lets the client send the first packet through the normal scheduling path and then directs all subsequent ones to the assigned worker based on the scheduling metadata returned in the first response.
}

\rev{
\subsection{Threats to Validity and Future Work}
\label{sec:scalability}
\label{sec:threats}

\textbf{Scalability.}
Worker scaling will not be expensive: each new worker consumes only lightweight pipeline metadata (\emph{e.g.}, QPN). The binding constraint actually comes from the client side, as more clients can generate more outstanding requests that need to share the 128\,K-entry on-switch task buffer. Scaling beyond this single-rack budget calls for a multi-rack extension, \emph{e.g.}, hierarchical scheduling where per-rack \sysname{} instances serve as local schedulers coordinated by a lightweight global layer. We leave this as future work.

\textbf{Switch-side overflow and overload control.}
Switch queues are inherently shallower than host-side queues, so pull-based in-network schedulers, including \sysname{}, are more vulnerable to overflow than push-based designs under open-loop arrivals. Our current safeguard is a global admission-control cap~(\S\ref{sec:implementation}). More sophisticated overload control is left as future work.

\textbf{Pipeline resource budget.}
Our Tofino prototype uses 11 of 12 pipeline stages. If other data-plane programs (\textit{e.g.}, load balancing, telemetry, or access control) need to be co-deployed, the pipeline layout has to be carefully planned to stay within the stage and SRAM limits.
}

\section{Related Work}
This section highlights the studies that are relevant to \sysname{} but have not been commented yet.

\textbf{Intra-server task scheduling.}
Early server-level systems~\cite{belay2014ix} often used hash-based techniques such as receiver-side scaling (RSS) to distribute requests across CPU cores. Beyond these randomized mechanisms, subsequent studies~\cite{kaffes2019shinjuku,didona2019size,iyer2023achieving,demoulin2021idling} explored centralized software schedulers that can make global dispatch decisions to improve core utilization and reduce tail latency. Complementary efforts leveraged hardware acceleration, including multicore system-on-a-chip designs~\cite{humphries2019mind}, FPGA-based NIC schedulers~\cite{lin2023ringleader}, and SmartNIC-assisted architectures~\cite{daglis2019rpcvalet}. These approaches focused on optimizing task distribution within a single server and are therefore orthogonal and complementary to the in-network scheduling problem addressed in this work.

\textbf{CPU scheduling.}
A broad class of systems dynamically adjusted CPU allocations in response to workload variations or interference, typically via user-level runtime operations that scale processing capacity on demand~\cite{fried2020caladan,ousterhout2019shenango,marty2019snap,belay2016ix,iorgulescu2018perfiso,qin2018arachne,kaufmann2019tas}. These mechanisms primarily regulated CPU resources directly. In contrast, our adaptive scheduling reacts to tail-latency behavior and modulates worker queue depth, providing a complementary control channel alongside CPU-level schedulers.

\textbf{RDMA in programmable switches.}
Integrating RDMA with programmable switches has expanded the design opportunity for high-performance distributed systems. Several efforts have demonstrated Tofino-based data planes issuing RDMA \texttt{READ/WRITE} requests directly~\cite{chen2023cowbird,jasny2024zero,kim2020tea,scazzariello2023high,sapio2021scaling}, enabling switches to actively access remote memory. Programmable switches have also been used to extend or optimize RDMA fabric capabilities~\cite{song2023network,li2024cepheus,ma2025rdmax}.

\section{Conclusion}

This paper presented \sysname{}, an RDMA-assisted in-network scheduler built on a programmable switch for microsecond-scale, latency-critical workloads. \sysname{} integrates bidirectional on-switch queuing, a switch-driven RDMA engine, and slice-aware scheduling to approximate centralized JBSQ behavior while efficiently supporting large tasks. Our study further revealed that real-world systems can diverge from theoretical predictions: shallow worker queues do not always improve tail latency. Leveraging this insight, \sysname{} incorporates an adaptive scheduler that dynamically tunes worker queue depths and worker-to-slice mappings, preserving work conservation and sustaining SLO-level tail latency across diverse workloads, and thus outperforms the SOTAs.

\begin{acks}
We would like to thank the anonymous reviewers and our shepherd, Gyuyeong Kim, for providing valuable feedback. This work was supported by the National Key R\&D Program of China under Grant 2023YFB2903903.
\end{acks}

\bibliographystyle{ACM-Reference-Format}
\bibliography{references}

\end{document}